\documentclass[sigconf,anonymous=false,screen=true,natbib=false]{acmart}
\RequirePackage[%
	datamodel=acmdatamodel,
	style=acmnumeric,
	backend=biber,
	sorting=none
]{biblatex}
\usepackage{amsmath,amsfonts}
\usepackage{graphicx}
\usepackage{booktabs}
\usepackage{xcolor}
\usepackage{tikz}
\usetikzlibrary{shapes.geometric}
\usetikzlibrary{tikzmark}
\usepackage{multicol}
\usepackage{enumitem}
\usepackage{calc}
\usepackage{todonotes}
\usepackage{siunitx}
\usepackage[htt]{hyphenat}
\usepackage{xurl}
\hypersetup{breaklinks=true}

\DeclareSIUnit\flop{\textsc{Fl}\textsc{Op}}
\DeclareSIUnit\cycle{\textsc{Cycle}}
\DeclareSIUnit[per-mode=symbol]\floppersec{\flop\per\second}
\DeclareSIQualifier{\doubleprecision}{FP64}
\DeclareSIQualifier{\singleprecision}{FP32}
\DeclareSIQualifier{\halfprecision}{FP16}

\colorlet{ok}{ACMGreen}
\colorlet{fullok}{ok}
\colorlet{prime}{ACMGreen!70!black}
\colorlet{indirectok}{ACMGreen!60!ACMOrange}
\colorlet{nonvendorok}{ACMGreen!80!white}
\colorlet{prettyok}{ACMOrange!90!white}
\colorlet{somesupport}{ACMOrange!50!ACMRed}
\colorlet{nope}{ACMRed}
\newcommand{\C}{C++}
\newcommand{\F}{Fortran}
\newcommand{\Python}{Python}
\newcommand{\mycirc}[1]{\tikz[inner sep=0pt, outer sep=0pt]{\node (X) {} [fill=#1, prime] (0, 0) circle [radius=1ex];}}

\newcommand{\fullok}{\mycirc{ok}}
\newcommand{\prettyok}{\tikz[inner sep=0pt, outer sep=0pt]{\node (X) [regular polygon, regular polygon sides=4, fill=prettyok, minimum size=2.8ex] at (0,0) {};}}
\newcommand{\nonvendorok}{\tikz[inner sep=0pt, outer sep=0pt]{\node (X) [regular polygon, regular polygon sides=3, fill=nonvendorok, minimum size=2.8ex] at (0,0) {};}}
\newcommand{\indirectok}{\tikz[inner sep=0pt, outer sep=0pt]{\node (X) [shape= semicircle, fill=indirectok, minimum size=1.33ex, rotate=180] at (0,0) {};}}
\newcommand{\somesupport}{\tikz[inner sep=0pt, outer sep=0pt]{\node (X) [star, star point height=0.8ex, fill=somesupport, minimum size=2.8ex] at (0,0) {};}}
\newcommand{\nope}{\tikz[inner sep=0pt, outer sep=0pt]{\path (X) [draw=nope, thick] (0,0) -- ++(1.4ex,1.4ex);}}

\newcommand{\refdesc}[1]{#1-desc}
\newcounter{pmdesc}
\makeatletter
\newcommand{\iflabelexists}[3]{\@ifundefined{myr@#1}{#3}{#2}}
\NewDocumentCommand{\refwithstate}{ m }{%
    \iflabelexists{#1}{}{%
        \global\expandafter\let\csname myr@#1\endcsname\@empty
        \refstepcounter{pmdesc}\label{#1}%
    }%
    \textsuperscript{%
    	\hyperlink{#1-desc}{\ref*{#1}}%
    }%
}
\makeatother
\newcommand{\lookup}[1]{%
	\csname#1\endcsname
}
\newcommand{\mymarker}[5]{%
	\colorlet{thiscolor}{black}%
	\tikz[baseline=(X.base)]{\node (X) [fill=thiscolor, text=white] {\hypertarget{#1-desc}{\hypersetup{hidelinks}\ref{#1}}
};} \textbf{#2 \textbullet{} #3 \textbullet{} \lookup{#4}:}%
}
\newcommand{\mycite}[1]{\autocite{#1}}

\newcommand{\catname}[1]{\emph{Category Name: \textbf{#1}}}

\title[Many Cores, Many Models]{Many Cores, Many Models: GPU Programming Model vs. Vendor Compatibility Overview}
\author{Andreas Herten}
\orcid{0000-0002-7150-2505}
\affiliation{%
	\institution{Forschungszentrum Jülich}
	\department{Jülich Supercomputing Centre}
	\city{Jülich}
	\country{Germany}%
}
\email{a.herten@fz-juelich.de}

\copyrightyear{2023} 
\acmYear{2023} 
\setcopyright{acmlicensed}\acmConference[SC-W 2023]{Workshops of The International Conference on High Performance Computing, Network, Storage, and Analysis}{November 12--17, 2023}{Denver, CO, USA}
\acmBooktitle{Workshops of The International Conference on High Performance Computing, Network, Storage, and Analysis (SC-W 2023), November 12--17, 2023, Denver, CO, USA}
\acmPrice{15.00}
\acmDOI{10.1145/3624062.3624178}
\acmISBN{979-8-4007-0785-8/23/11}



\begin{abstract}
	In recent history, GPUs became a key driver of compute performance in HPC. With the installation of the Frontier supercomputer, they became the enablers of the Exascale era; further largest-scale installations are in progress (Aurora, El Capitan, JUPITER). But the early-day dominance by NVIDIA and their CUDA programming model has changed: The current HPC GPU landscape features three vendors (AMD, Intel, NVIDIA), each with native and derived programming models. The choices are ample, but not all models are supported on all platforms, especially if support for Fortran is needed; in addition, some restrictions might apply. It is hard for scientific programmers to navigate this abundance of choices and limits.	This paper gives a guide by matching the GPU platforms with supported programming models, presented in a concise table and further elaborated in detailed comments. An assessment is made regarding the level of support of a model on a platform.
\end{abstract}

\DeclareFieldFormat{titlecase}{#1}

\begin{document}

\maketitle

\section{Introduction}
Taking the TOP500 list of June 2023 as a reference~\autocite{top500-23-06}, more than \qty{60}{\percent} of the available \unit{\floppersec} are delivered by Graphics Processing Units (GPUs). The devices were first installed in HPC systems in the mid 2000s and steadily matured over the next decades. The most-recent culmination came in 2022, when the first Exascale supercomputer, Frontier at Oak Ridge National Lab, was added to the TOP500 list, utilizing more than \num{37000} GPUs to deliver \qty{1194}{\peta\floppersec} (\emph{Rmax}) of compute performance -- alone delivering about \qty{20}{\percent} of the entire list's compute performance. Further largest-scale installations using GPUs are planned or already on the way, like Aurora (at Argonne National Lab), El Capitan (at Lawrence Livermore National Lab), or JUPITER (at Jülich Supercomputing Centre).

While the first years of GPU usage in HPC was dominated by NVIDIA GPUs and NVIDIA's CUDA programming model, the landscape significantly changed in the last years. Frontier utilizes AMD GPUs ($\num{37 888} \times$ AMD Radeon Instinct MI250X) and Aurora uses Intel GPUs ($\num{63744} \times$ 
Intel Data Center GPU Max Series, codename \emph{Ponte Vecchio}); also El Capitan will use next-generation AMD GPUs (AMD Radeon Instinct MI300A). Each GPU platform has a selected major native programming model: CUDA for NVIDIA, HIP for AMD, and SYCL for Intel\footnote{Intel bundles their parallel programming infrastructure into \emph{oneAPI}, which includes -- amongst others -- DPC++, their SYCL implementation. Next to SYCL, also OpenMP is a prominently promoted programming model by Intel.}. They are augmented with further vendor- or community-driven models, usually presenting higher-level abstractions. Examples are OpenMP and OpenACC as the two major directive-based models; Kokkos, RAJA, and Alpaka which enable GPU programming through high-level abstractions for parallel algorithms and data management; and Standard-based parallelism which utilizes modern features of programming languages themselves to access GPUs. The key scientific programming language is C++ (sometimes programmed in a plain C-style), but also Fortran is still prevalent in many scientific applications. Also Python has become a popular choice in recent years~\autocite{tiobe,pypl}; as an even higher-level, interpreted programming language it relies on \emph{backends} in lower-level languages -- mostly C/C++ -- and rather implements interfaces.

Although the evolving combinatorial explosion of choices\footnote{GPU platforms $\times$ programming models $\times$ programming languages} is a good sign for the health of the GPU ecosystem, the field can at times be hard to navigate -- for established GPU developers but especially for novice users. With the selection made in this paper, more than 50 routes for programming a GPU device are identified when no further limitations (pre-)exist. This work gives a guide into the current GPU programming ecosystem, by categorizing the individual possibilities in a concise table and explaining each combination in detail.


The main contributions of this paper are the categories of rating support of programming models on GPU devices in \autoref{sec:ratings}, the application in the overview table in \autoref{table:compat}, and the accompanying list of explanations in \autoref{sec:descriptions}, with many links to further resources.

The paper is structured as follows: In ~\autoref{sec:ratings}, the six rating categories are explained in detail and some comments to the method are made. In \autoref{sec:descriptions}, the core of this paper, the overview table (\autoref{table:compat}), is presented and explained with detailed comments for each possible choice. {\renewcommand{\sectionautorefname}{Section}\autoref{sec:limitations}} shows limitations and caveats of the table and methodology. Finally, \autoref{sec:conclusion} concludes this paper.

\section{Related Work}

While no other work is known outlining and assessing the usage of programming models on certain GPU devices to the extent presented here, related work previously compared specific aspects or sub-sets, usually with a focus on performance. \citeauthor{babelstreamdoconcurrent}~\autocite{babelstreamdoconcurrent} compared performance for standard language parallelism in Fortran, by using the BabelStream benchmarks. \citeauthor{benchmarkinglumi}~\autocite{benchmarkinglumi} cast a wider net around applications, but focused solely on the LUMI supercomputer. \citeauthor{hammondgtc} evaluated several NVIDIA-compatible GPU programming models in~\autocite{hammondgtc}. A very detailed comparison of GPU support through various OpenMP-capable compilers was given at the 2022 ECP Community BOF Days~\autocite{ecpopenmpbof}. Deeper, more technical insights can be gained by dedicated validation suites~\autocite{openmpvvproceedings,openaccvvproceedings}. Further examples are discussed in \autoref{sec:descriptions}.

\section{Method, Categories}
\label{sec:ratings}

\begin{figure*}[t]
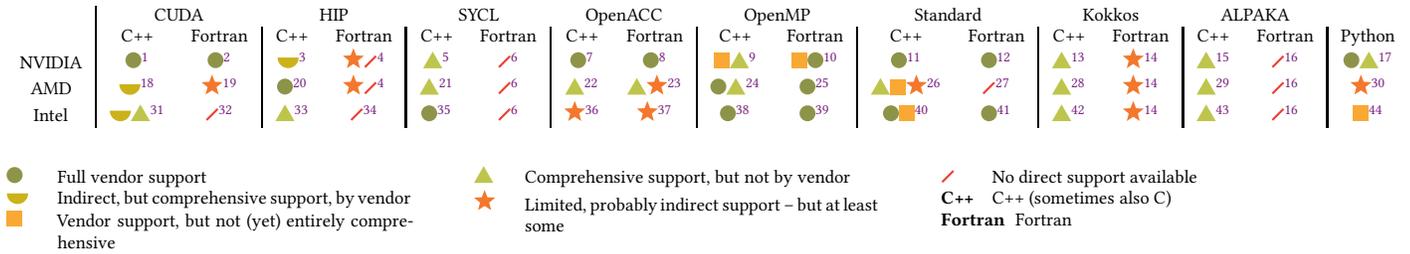

	{%
	\footnotesize
	\begin{tabular}{c|cc|cc|cc|cc|cc|cc|cc|cc|c}
  &
        \multicolumn{2}{c}{CUDA} &
        \multicolumn{2}{c}{HIP} &
        \multicolumn{2}{c}{SYCL} &
        \multicolumn{2}{c}{OpenACC} &
        \multicolumn{2}{c}{OpenMP} &
        \multicolumn{2}{c}{Standard} &
        \multicolumn{2}{c}{Kokkos} &
        \multicolumn{2}{c}{ALPAKA} &
        \\
  &
        \C &\F &
        \C &\F &
        \C &\F &
        \C &\F &
        \C &\F &
        \C &\F &
        \C &\F &
        \C &\F &
        \Python 
        \\
  NVIDIA &
  \fullok\refwithstate{cudac} & \fullok\refwithstate{cudafortran} & \indirectok\refwithstate{nvidiahip} & \somesupport\nope\refwithstate{nvidiahipfortran} & \nonvendorok\refwithstate{nvidiasycl} & \nope\refwithstate{syclfortran} & \fullok\refwithstate{openaccc} & \fullok\refwithstate{openaccfortran} & \prettyok\nonvendorok\refwithstate{nvidiaopenmpc} & \prettyok\fullok\refwithstate{nvidiaopenmpfortran} & \fullok\refwithstate{nvidiastandardc} & \fullok\refwithstate{nvidiastandardfortran} & \nonvendorok\refwithstate{nvidiakokkosc} & \somesupport\refwithstate{nvidiakokkosfortran} & \nonvendorok\refwithstate{nvidiaalpakac} & \nope\refwithstate{nvidiaalpakafortran} & \fullok\nonvendorok\refwithstate{nvidiapython} \\
  AMD &
  \indirectok\refwithstate{amdcudac} & \somesupport\refwithstate{amdcudafortran} & \fullok\refwithstate{amdhipc} & \somesupport\nope\refwithstate{nvidiahipfortran} & \nonvendorok\refwithstate{amdsyclc} & \nope\refwithstate{syclfortran} & \nonvendorok\refwithstate{amdopenaccc} & \nonvendorok\somesupport\refwithstate{amdopenaccfortran} & \fullok\nonvendorok\refwithstate{amdopenmpc} & \fullok\refwithstate{amdopenmpfortran} & \nonvendorok\prettyok\somesupport\refwithstate{amdstandardc} & \nope\refwithstate{amdstandardfortran} & \nonvendorok\refwithstate{amdkokkosc} & \somesupport\refwithstate{nvidiakokkosfortran} & \nonvendorok\refwithstate{amdalpakac} & \nope\refwithstate{nvidiaalpakafortran} & \somesupport\refwithstate{amdpython} \\
  Intel &
  \indirectok\nonvendorok\refwithstate{intelcudac} & \nope\refwithstate{intelcudafortran} & \nonvendorok\refwithstate{intelhipc} & \nope\refwithstate{intelhipfortran} & \fullok\refwithstate{intelsyclc} & \nope\refwithstate{syclfortran} & \somesupport\refwithstate{intelopenaccc} & \somesupport\refwithstate{intelopenaccfortran} & \fullok\refwithstate{intelopenmpc} & \fullok\refwithstate{intelopenmpfortran} & \fullok\prettyok\refwithstate{intelstandardc} & \fullok\refwithstate{intelstandardfortran} & \nonvendorok\refwithstate{intelkokkosc} & \somesupport\refwithstate{nvidiakokkosfortran} & \nonvendorok\refwithstate{intelalpakac} & \nope\refwithstate{nvidiaalpakafortran} & \prettyok\refwithstate{intelpython} \\
  \end{tabular}
	\begin{multicols}{3}
	    \begin{description}[leftmargin=\widthof{C++\quad}, style=sameline]
	        \item[\fullok] Full vendor support
	        \item[\indirectok] Indirect, but comprehensive support, by vendor
	        \item[\prettyok] Vendor support, but not (yet) entirely comprehensive
	        \item[\nonvendorok] Comprehensive support, but not by vendor
	        \item[\somesupport] Limited, probably indirect support -- but at least some
	        \item[\nope] No direct support available
	        \item[\C] C++ (sometimes also C)
	        \item[\F] Fortran
		\end{description}
	\end{multicols}%
	}\vspace*{-4ex}%
	\caption{Overview table comparing a selection of major GPU programming models with the current state of support by the three vendors of dedicated HPC GPUs (AMD, Intel, NVIDIA) for \C{} and \F{}. See \autoref{desc:categories} for more detailed explanations of the categories.}%
	\label{table:compat}
\end{figure*}

\autoref{table:compat} matches the three GPU vendors AMD, Intel, and NVIDIA (row) with programming models (columns). Each column is additionally separated into two sub-columns for the two programming language \C{} and \F{}. The presented programming models are the three \emph{native} models (CUDA, HIP, SYCL), the two major directive-based models (OpenMP, OpenACC), two examples of community-driven, higher-level models with foci on platform-portability (Kokkos, Alpaka), and the upcoming GPU usage through standard features in the programming language (\emph{Standard}). In addition, support by Python is summarized for each platform. In total, 51 possible combinations are explored and explained in 44 unique descriptions. 

No restrictions are exposed regarding language version of \C{} and \F{}, as it would add another level of complexity and is usually not a limiting factor for scientists due to backward compatibility. While \C{} is required by most programming models, some models can also be used in a C-like manner. For the sake of brevity, this paper considers \C{}. 

To assess and describe the 51 possible combinations, a review of available literature and online resources was conducted. The available information and its level of detail and coverage varies significantly between the models; it is most extensive for the native models and sparsest for some community-driven, explorative implementations. The combinations are assessed by this available information and -- to a limited extend -- the experience of the author. The paper strives to be objective and derives the rating with thorough descriptions. Of course, classifying into six distinct categories has some limitations, outlined within the descriptions themselves and discussed further in \autoref{sec:limitations}.

This work introduces six categories to assess the coverage of a certain combination of programming model and language on a respective GPU platform. The categories are indicated by symbols, reaching from \fullok{} (full support) to \nope{} (no support), with various intermediate steps. The following list explains the categorizing symbols and also names the categories for completeness.

\begin{description}[leftmargin=1.6em, style=sameline]
\label{desc:categories}
	\item[\raisebox{0pt}{\fullok}] The programming model for this language is fully supported on this GPU platform by the vendor. The vendor provides complete implementation of the combination and extensive documentation. The model is regularly updated and the vendor provides support in case of errors. \catname{full support}
	\item[\raisebox{0pt}{\indirectok}] The combination of programming model and language is indirectly, but comprehensively supported by the vendor of the GPU device. This is usually achieved by (semi-)automatically mapping/translating a \emph{foreign} model to a \emph{native} one. \catname{indirect good support}
	\item[\raisebox{0pt}{\prettyok}] Model/language are supported on this GPU device by the vendor, but the support is not (yet) comprehensive. Usually, the model can be used for the majority of applications, but some specific features are not available. \catname{some support}
	\item[\raisebox{0pt}{\nonvendorok}] Comprehensive support is available for this combination of programming model and programming language on a GPU device, but not by the vendor of the GPU device itself. Usually, higher-level models driven by the community implement support and utilize vendor-native infrastructure in the background, unexposed to the user. \catname{non-vendor good support}
	\item[\raisebox{0pt}{\somesupport}] Some very limited support is available for this programming model and language on a certain GPU device. The support might be indirect, through extensive effort by the user, and/or very incomplete. \catname{limited support}
	\item[\raisebox{0pt}{\nope}] No direct support is available for the model/language on the device. There are certainly ways to still utilize the device, like creating custom headers and linking to libraries directly, or utilizing \texttt{ISO\_C\_BINDING} in Fortran. \catname{no support}
\end{description}

The following \autoref{sec:descriptions} lists the descriptions of each possible combination, referring the categorizing symbols of \autoref{table:compat} with reference numbers\footnote{In the PDF version of this document, both number can be clicked and move between table and description.}. In each description, links to online resources are overlaid as hyperlinks. The key references for an item are included as entries in the bibliography. The three native programming models (CUDA, HIP, SYCL) are explained in greatest details for their main platform (NVIDIA, AMD, Intel GPU devices, respectively). At times, descriptions for entries are similar. This is by choice due to the encyclopedic nature of the document in which readers might look up only single entries.

\section{Descriptions}
\label{sec:descriptions}

{\setlength{\parindent}{0pt}%
\setlength{\parskip}{0.66ex}%
\mymarker{cudac}{NVIDIA}{CUDA}{C}{fullok} CUDA C/C++ is supported on NVIDIA GPUs through the \href{https://developer.nvidia.com/cuda-toolkit}{CUDA Toolkit}. First released in 2007, the toolkit covers nearly all aspects of the NVIDIA platform: an API for programming (incl. language extensions), libraries, tools for profiling and debugging, compiler, management tools, and more. The current version is CUDA 12.2. Usually, when referring to \emph{CUDA} without any additional context, the CUDA API is meant. While incorporating some Open Source components, the CUDA platform in its entirety is proprietary and closed sourced. The low-level CUDA instruction set architecture is PTX, to which higher languages like the CUDA C/C++ are translated to. PTX is compiled to SASS, the binary code executed on the device. As it is the reference for platform, the support for NVIDIA GPUs through CUDA C/C++ is very comprehensive. In addition to support through the CUDA toolkit, NVIDIA GPUs can also be \href{https://llvm.org/docs/CompileCudaWithLLVM.html}{used by Clang}, utilizing the LLVM toolchain to emit PTX code and compile it subsequently.  \mycite{CUDA} 

\mymarker{cudafortran}{NVIDIA}{CUDA}{F}{fullok} CUDA Fortran, a proprietary Fortran extension by NVIDIA, is supported on NVIDIA GPUs via the \href{https://developer.nvidia.com/hpc-sdk}{NVIDIA HPC SDK} (\emph{NVHPC}). NVHPC implements most features of the CUDA API in Fortran and is activated through the \texttt{-cuda} switch in the \texttt{nvfortran} compiler. The CUDA extensions for Fortran are modeled closely after the CUDA C/C++ definitions. In addition to creating explicit kernels in Fortran, CUDA Fortran also supports \emph{cuf kernels}, a way to let the compiler generate GPU parallel code automatically. Very recently, \href{https://reviews.llvm.org/D150159}{CUDA Fortran support was also merged into Flang}, the LLVM-based Fortran compiler.  \mycite{CUDAFortran} 

\mymarker{nvidiahip}{NVIDIA}{HIP}{C}{indirectok} \href{https://github.com/ROCm-Developer-Tools/HIP}{HIP} programs can directly use NVIDIA GPUs via a CUDA backend. As HIP is strongly inspired by CUDA, the mapping is relatively straight-forward; API calls are named similarly (for example: \texttt{hipMalloc()} instead of \texttt{cudaMalloc()}) and keywords of the kernel syntax are identical. HIP also supports some CUDA libraries and creates interfaces to them (like \texttt{hipblasSaxpy()} instead of \texttt{cublasSaxpy()}). To target NVIDIA GPUs through the HIP compiler (\texttt{hipcc}), \texttt{HIP\_PLATFORM=nvidia} needs to be set in the environment. In order to initially create a HIP code from CUDA, AMD offers the \href{https://github.com/ROCm-Developer-Tools/HIPIFY}{HIPIFY} conversion tool.  \mycite{HIP} 

\mymarker{nvidiahipfortran}{NVIDIA, AMD}{HIP}{F}{['somesupport', 'nope']} No Fortran version of HIP exists; HIP is solely a C/C++ model. But AMD offers an extensive set of ready-made interfaces to the HIP API and HIP and ROCm libraries with \href{https://github.com/ROCmSoftwarePlatform/hipfort}{hipfort} (MIT-licensed). All interfaces implement C functionality and CUDA-like Fortran extensions, for example to write kernels, are available.  \mycite{hipfort} 

\mymarker{nvidiasycl}{NVIDIA}{SYCL}{C}{nonvendorok} No direct support for \href{https://www.khronos.org/sycl/}{SYCL} is available by NVIDIA, but SYCL can be used on NVIDIA GPUs through multiple venues. First, SYCL can be \href{https://github.com/intel/llvm/blob/sycl/sycl/doc/GetStartedGuide.md\#build-dpc-toolchain-with-support-for-nvidia-cuda}{used through DPC++}, an Open-Source LLVM-based compiler project \href{https://github.com/intel/llvm}{led by Intel}. The DPC++ infrastructure is also available through Intel\textquotesingle s commercial \href{https://www.intel.com/content/www/us/en/developer/tools/oneapi/dpc-compiler.html}{oneAPI toolkit} (\emph{Intel oneAPI DPC++/C++}) as \href{https://developer.codeplay.com/products/oneapi/nvidia/2023.2.1/guides/get-started-guide-nvidia}{a dedicated plugin}. Upstreaming SYCL support directly into LLVM is an \href{https://github.com/intel/llvm/issues/49}{ongoing effort}, which started \href{https://lists.llvm.org/pipermail/cfe-dev/2019-January/060811.html}{in 2019}. Further, SYCL can be used via \href{https://github.com/OpenSYCL/OpenSYCL/}{Open SYCL} (previously called hipSYCL), an independently developed SYCL implementation, using NVIDIA GPUs either through the CUDA support of LLVM or the \texttt{nvc++} compiler of NVHPC. A third popular possibility was the NVIDIA GPU support in \href{https://github.com/codeplaysoftware/sycl-for-cuda/tree/cuda}{ComputeCpp of CodePlay}; though \href{https://developer.codeplay.com/products/computecpp/ce/home/}{the product became unsupported in September 2023}. In case LLVM is involved, SYCL implementations can rely on CUDA support in LLVM, which needs the CUDA toolkit available for the final compilations parts beyond PTX. In order to translate a CUDA code to SYCL, Intel offers the \href{https://github.com/oneapi-src/SYCLomatic}{SYCLomatic} conversion tool.  \mycite{intelllvm,opensyclproceedings} 

\mymarker{syclfortran}{NVIDIA, AMD, Intel}{SYCL}{F}{nope} SYCL is a C++-based programming model (C++17) and by its nature does not support Fortran. Also, no pre-made bindings are available.  \mycite{khronossycl} 

\mymarker{openaccc}{NVIDIA}{OpenACC}{C}{fullok} OpenACC C/C++ on NVIDIA GPUs is supported most extensively through the \href{https://developer.nvidia.com/hpc-sdk}{NVIDIA HPC SDK}. Beyond the bundled libraries, frameworks, and other models, the NVIDIA HPC SDK also features the \texttt{nvc}/\texttt{nvc++} compilers, in which \href{https://docs.nvidia.com/hpc-sdk/compilers/hpc-compilers-user-guide/index.html\#acc-use}{OpenACC support} can be enabled with the \texttt{-acc\ -gpu}. The support of OpenACC in this vendor-delivered compiler is very comprehensive, it conforms to version 2.7 of the specification. A variety of compile options are available to modify the compilation process. In addition to NVIDIA HPC SDK, good support is also available in GCC since GCC 5.0, \href{https://gcc.gnu.org/wiki/OpenACC}{supporting OpenACC 2.6} through the \texttt{nvptx} architecture. The compiler switch to enable OpenACC in \texttt{gcc}/\texttt{g++} is \texttt{-fopenacc}, further options are available. Further, the \href{https://csmd.ornl.gov/project/clacc}{Clacc compiler} implements OpenACC support into the LLVM toolchain, adapting the Clang frontend. As a central design aspect, it translates OpenACC to OpenMP as part of the compilation process. OpenACC can be activated in a Clacc-\texttt{clang} via \texttt{-fopenacc}, and further compiler options exist, mostly leveraging OpenMP options. A recent study by \href{https://ieeexplore.ieee.org/document/10029456}{Jarmusch et al.} compared these compilers for coverage of the OpenACC 3.0 specification.  \mycite{nvhpc,gccopenacc,claccieee,jarmusch22} 

\mymarker{openaccfortran}{NVIDIA}{OpenACC}{F}{fullok} Support of OpenACC Fortran on NVIDIA GPUs is similar to OpenACC C/C++, albeit not identical. First, \href{https://developer.nvidia.com/hpc-sdk}{NVIDIA HPC SDK} supports OpenACC in Fortran through the included \texttt{nvfortran} compiler, with options like for the C/C++ compilers. In addition, also \href{https://gcc.gnu.org/wiki/OpenACC}{GCC supports OpenACC} through the \texttt{gfortran} compiler with identical compiler options to the C/C++ compilers. Further, similar to OpenACC support in LLVM for C/C++ through \emph{Clacc} contributions, the LLVM frontend for Fortran, \href{https://flang.llvm.org/docs/}{Flang} (the successor of \emph{F18}, not \emph{classic Flang}), \href{https://flang.llvm.org/docs/OpenACC.html}{supports OpenACC} as well. Support was initially contributed through the \href{https://ieeexplore.ieee.org/document/9651310}{Flacc project} and now resides in the main LLVM project. Finally, the \href{https://www.hpe.com/psnow/doc/a50002303enw}{HPE Cray Programming Environment} supports \href{https://cpe.ext.hpe.com/docs/cce/man7/intro_openacc.7.html}{OpenACC Fortran}; in \texttt{ftn-hacc}.  \mycite{nvhpc,gccopenacc,flaccieee} 

\mymarker{nvidiaopenmpc}{NVIDIA}{OpenMP}{C}{['prettyok', 'nonvendorok']} OpenMP in C/C++ is supported on NVIDIA GPUs (\emph{Offloading}) through multiple venues, similarly to OpenACC. First, the NVIDIA HPC SDK supports \href{https://docs.nvidia.com/hpc-sdk/compilers/hpc-compilers-user-guide/index.html\#openmp-use}{OpenMP GPU offloading} in both \texttt{nvc} and \texttt{nvc++}, albeit only a subset of the entire OpenMP 5.0 standard (see \href{https://docs.nvidia.com/hpc-sdk/compilers/hpc-compilers-user-guide/index.html\#openmp-subset}{the documentation for supported/unsupported features}). The key compiler option is \texttt{-mp}. Also in GCC, \href{https://gcc.gnu.org/wiki/Offloading}{OpenMP offloading} can be used to NVIDIA GPUs; the compiler switch is \texttt{-fopenmp}, with options delivered through \href{https://gcc.gnu.org/onlinedocs/gcc/C-Dialect-Options.html\#index-foffload}{\texttt{-foffload} and \texttt{-foffload-options}}. GCC \href{https://gcc.gnu.org/onlinedocs/gcc-13.1.0/libgomp/OpenMP-Implementation-Status.html}{currently supports OpenMP 4.5 entirely}, while OpenMP features of 5.0, 5.1, and, 5.2 are currently being implemented. Similarly in Clang, where \href{https://clang.llvm.org/docs/OffloadingDesign.html}{OpenMP offloading to NVIDIA GPUs} is supported and enabled through \texttt{-fopenmp\ -fopenmp-targets=nvptx64}, with offload architectures selected via \texttt{-\/-offload-arch=native} (or similar). Clang implements \href{https://clang.llvm.org/docs/OpenMPSupport.html\#openmp-implementation-details}{nearly all OpenMP 5.0 features and most of OpenMP 5.1/5.2}. In the HPE Cray Programming Environment, a \href{https://cpe.ext.hpe.com/docs/cce/man7/intro_openmp.7.html}{subset of OpenMP 5.0/5.1 is supported} for NVIDIA GPUs. It can be activated through \texttt{-fopenmp}. Also \href{https://github.com/ROCm-Developer-Tools/aomp/}{AOMP}, AMD\textquotesingle s Clang/LLVM-based compiler, supports NVIDIA GPUs. Support of OpenMP features in the compilers was recently discussed in the \href{https://www.openmp.org/wp-content/uploads/2022_ECP_Community_BoF_Days-OpenMP_RoadMap_BoF.pdf}{OpenMP ECP BoF 2022}.  \mycite{nvhpc,gccopenmp,clangopenmp,hpepe} 

\mymarker{nvidiaopenmpfortran}{NVIDIA}{OpenMP}{F}{['prettyok', 'fullok']} OpenMP in Fortran is supported on NVIDIA GPUs nearly identical to C/C++. \href{https://docs.nvidia.com/hpc-sdk/compilers/hpc-compilers-user-guide/index.html\#openmp-use}{NVIDIA HPC SDK\textquotesingle s \texttt{nvfortran}} implements support, \href{https://gcc.gnu.org/wiki/openmp}{GCC\textquotesingle s \texttt{gfortran}}, \href{https://flang.llvm.org/docs/}{LLVM\textquotesingle s Flang} (through \texttt{-mp}, and \href{https://flang.llvm.org/docs/GettingStarted.html\#openmp-target-offload-build}{only when Flang is compiled via Clang}), and also the \href{https://cpe.ext.hpe.com/docs/cce/man7/intro_openmp.7.html}{HPE Cray Programming Environment}.  \mycite{nvhpc,gccopenmp,hpepe,flang} 

\mymarker{nvidiastandardc}{NVIDIA}{Standard}{C}{fullok} Standard language parallelism of C++, namely algorithms and data structures of the \emph{parallel STL}, is supported on NVIDIA GPUs \href{https://docs.nvidia.com/hpc-sdk/compilers/c++-parallel-algorithms/index.html}{through the \texttt{nvc++} compiler of the NVIDIA HPC SDK}. The key compiler option is \texttt{-stdpar=gpu}, which enables offloading of parallel algorithms to the GPU. Also, currently Open SYCL \href{https://github.com/OpenSYCL/OpenSYCL/pull/1088}{is in the process of implementing support for pSTL algorithms}, enabled via \texttt{-\/-hipsycl-stdpar}. Further, \href{https://intel.github.io/llvm-docs/GetStartedGuide.html\#build-dpc-toolchain-with-support-for-nvidia-cuda}{NVIDIA GPUs can be targeted from Intel\textquotesingle s DPC++ compiler}, enabling usage of pSTL algorithms implemented in Intel\textquotesingle s Open Source \href{https://github.com/oneapi-src/oneDPL}{oneDPL} (\emph{oneAPI DPC++ Library}) on NVIDIA GPUs. Finally, a \href{https://discourse.llvm.org/t/rfc-openmp-offloading-backend-for-c-parallel-algorithms/73468}{current proposal in the LLVM community} aims at implementing pSTL support through an OpenMP backend.  \mycite{nvhpc,opensyclproceedings,onedpl} 

\mymarker{nvidiastandardfortran}{NVIDIA}{Standard}{F}{fullok} Standard language parallelism of Fortran, mainly \texttt{do\ concurrent}, is supported on NVIDIA GPUs \href{https://developer.nvidia.com/blog/accelerating-fortran-do-concurrent-with-gpus-and-the-nvidia-hpc-sdk/}{through the \texttt{nvfortran} compiler of the NVIDIA HPC SDK}. As for the C++ case, it is enabled through the \texttt{-stdpar=gpu} compiler option.  \mycite{nvhpc} 

\mymarker{nvidiakokkosc}{NVIDIA}{Kokkos}{C}{nonvendorok} \href{https://github.com/kokkos/kokkos}{Kokkos} supports NVIDIA GPUs in C++. Kokkos has \href{https://kokkos.github.io/kokkos-core-wiki/requirements.html}{multiple backends} available with NVIDIA GPU support: a native CUDA C/C++ backend (using \texttt{nvcc}), an NVIDIA HPC SDK backend (using CUDA support in \texttt{nvc++}), and a Clang backend, using either Clang\textquotesingle s CUDA support directly or \href{https://docs.nersc.gov/development/programming-models/kokkos/}{via the OpenMP offloading facilities} (via \texttt{clang++}).  \mycite{kokkos} 

\mymarker{nvidiakokkosfortran}{NVIDIA, AMD, Intel}{Kokkos}{F}{somesupport} Kokkos is a C++ programming model, but an official compatibility layer for Fortran (\href{https://github.com/kokkos/kokkos-fortran-interop}{\emph{Fortran Language Compatibility Layer}, FLCL}) is available. Through this layer, GPUs can be used as supported by Kokkos C++.  \mycite{kokkos} 

\mymarker{nvidiaalpakac}{NVIDIA}{ALPAKA}{C}{nonvendorok} \href{https://github.com/alpaka-group/alpaka}{Alpaka} supports NVIDIA GPUs in C++ (C++17), either through the NVIDIA CUDA C/C++ compiler \texttt{nvcc} or LLVM/Clang\textquotesingle s support of CUDA in \texttt{clang++}.  \mycite{alpaka} 

\mymarker{nvidiaalpakafortran}{NVIDIA, AMD, Intel}{ALPAKA}{F}{nope} Alpaka is a C++ programming model and no ready-made Fortran support exists.  \mycite{alpaka} 

\mymarker{nvidiapython}{NVIDIA}{etc}{Python}{['fullok', 'nonvendorok']} Using NVIDIA GPUs from Python code can be achieved through multiple venues. NVIDIA itself offers \href{https://github.com/NVIDIA/cuda-python}{CUDA Python}, a package delivering low-level interfaces to CUDA C/C++. Typically, code is not directly written using CUDA Python, but rather CUDA Python functions as a backend for higher level models. CUDA Python is available on PyPI as \href{https://pypi.org/project/cuda-python/}{\texttt{cuda-python}}. An alternative to CUDA Python from the community is \href{https://github.com/inducer/pycuda}{PyCUDA}, which adds some higher-level features and functionality and comes with its own C++ base layer. PyCUDA is available on PyPI as \href{https://pypi.org/project/pycuda/}{\texttt{pycuda}}. The most well-known, higher-level abstraction is \href{https://cupy.dev/}{CuPy}, which implements primitives known from Numpy with GPU support, offers functionality for defining custom kernels, and bindings to libraries. CuPy is available on PyPI as \href{https://pypi.org/project/cupy-cuda12x/}{\texttt{cupy-cuda12x}} (for CUDA 12.x). Two packages arguably providing even higher abstractions are Numba and CuNumeric. \href{http://numba.pydata.org/}{Numba} offers access to NVIDIA GPUs and features acceleration of functions through Python decorators (\emph{functions wrapping functions}); it is available as \href{https://pypi.org/project/numba/}{\texttt{numba}} on PyPI. \href{https://github.com/nv-legate/cunumeric}{cuNumeric}, a project by NVIDIA, allows to access the GPU via Numpy-inspired functions (like CuPy), but utilizes the \href{https://github.com/nv-legate/legate.core}{Legate library} to transparently scale to multiple GPUs.  \mycite{cudapython,pycuda,cupy,numba,cunumeric} 

\mymarker{amdcudac}{AMD}{CUDA}{C}{indirectok} While CUDA is not directly supported on AMD GPUs, it can be translated to HIP through AMD\textquotesingle s \href{https://github.com/ROCm-Developer-Tools/HIPIFY}{HIPIFY}. Using \texttt{hipcc} and \texttt{HIP\_PLATFORM=amd} in the environment, CUDA-to-HIP-translated code can be executed.  \mycite{HIP} 

\mymarker{amdcudafortran}{AMD}{CUDA}{F}{somesupport} No direct support for CUDA Fortran on AMD GPUs is available, but AMD offers a source-to-source translator, \href{https://github.com/ROCmSoftwarePlatform/gpufort}{GPUFORT}, to convert some CUDA Fortran to either Fortran with OpenMP (via \href{https://github.com/ROCm-Developer-Tools/aomp}{AOMP}) or Fortran with HIP bindings and extracted C kernels (via \href{https://github.com/ROCmSoftwarePlatform/hipfort}{hipfort}). As stated in the project repository, the covered functionality is \href{https://github.com/ROCmSoftwarePlatform/gpufort\#limitations}{driven by use-case requirements}; the last commit is two years old.  \mycite{gpufort} 

\mymarker{amdhipc}{AMD}{HIP}{C}{fullok} \href{https://github.com/ROCm-Developer-Tools/HIP}{HIP} C++ is the \emph{native} programming model for AMD GPUs and, as such, fully supports the devices. It is part of AMD\textquotesingle s GPU-targeted \href{https://rocm.docs.amd.com/en/latest/}{ROCm platform}, which includes compilers, libraries, tool, and drivers and mostly consists of Open Source Software. HIP code can be compiled with \href{https://github.com/ROCm-Developer-Tools/HIPCC}{\texttt{hipcc}}, utilizing the correct environment variables (like \texttt{HIP\_PLATFORM=amd}) and compiler options (like \texttt{-\/-offload-arch=gfx90a}). \texttt{hipcc} is a \emph{compiler driver} (wrapper script) which assembles the correct compilation string, finally calling \href{https://github.com/RadeonOpenCompute/llvm-project}{AMD\textquotesingle s Clang compiler} to generate host/device code (using the \href{https://llvm.org/docs/AMDGPUUsage.html}{AMDGPU backend}).  \mycite{HIP} 

\mymarker{amdsyclc}{AMD}{SYCL}{C}{nonvendorok} No direct support for SYCL is available by AMD for their GPU devices. But like for the NVIDIA ecosystem, SYCL C++ can be used on AMD GPUs through third-party software. First, \href{https://github.com/OpenSYCL/OpenSYCL}{Open SYCL} (previously \emph{hipSYCL}) supports AMD GPUs, relying on HIP/ROCm support in Clang. All available \href{https://github.com/OpenSYCL/OpenSYCL/blob/develop/doc/compilation.md}{internal compilation models} can target AMD GPUs. Second, also AMD GPUs can be targeted through both \href{https://github.com/intel/llvm/blob/sycl/sycl/doc/GetStartedGuide.md\#build-dpc-toolchain-with-support-for-hip-amd}{DPC++}, Intel\textquotesingle s LLVM-based Open Source compiler, and the commercial version included in the \href{https://www.intel.com/content/www/us/en/developer/tools/oneapi/dpc-compiler.html}{oneAPI toolkit} (via an \href{https://developer.codeplay.com/products/oneapi/amd/2023.2.1/guides/get-started-guide-amd}{AMD ROCm plugin}). In comparison to SYCL support for CUDA, no conversion tool like SYCLomatic exists.  \mycite{opensyclproceedings,intelllvm} 

\mymarker{amdopenaccc}{AMD}{OpenACC}{C}{nonvendorok} OpenACC C/C++ is not supported by AMD itself, but third-party support is available for AMD GPUs through GCC or Clacc (similarly to their support of OpenACC C/C++ for NVIDI GPUS). In \href{https://gcc.gnu.org/wiki/Offloading}{GCC, OpenACC support} can be activated through \texttt{-fopenacc}, and further specified for AMD GPUs with, for example, \texttt{-foffload=amdgcn-amdhsa="-march=gfx906"}. \href{https://csmd.ornl.gov/project/clacc}{Clacc also supports OpenACC C/C++ on AMD GPUs} by translating OpenACC to OpenMP and using LLVM\textquotesingle s AMD support. The enabling compiler switch is \texttt{-fopenacc}, and AMD GPU targets can be further specified by, for example, \texttt{-fopenmp-targets=amdgcn-amd-amdhsa}. \href{/\%22https://github.com/intel/intel-application-migration-tool-for-openacc-to-openmp/\%22}{Intel\textquotesingle s OpenACC to OpenMP source-to-source translator} can also be used for AMD\textquotesingle s platform.  \mycite{gccopenacc,claccieee} 

\mymarker{amdopenaccfortran}{AMD}{OpenACC}{F}{['nonvendorok', 'somesupport']} No native support for OpenACC on AMD GPUs for Fortran is available, but AMD supplies \href{https://github.com/ROCmSoftwarePlatform/gpufort}{GPUFORT}, a research project to source-to-source translate OpenACC Fortran to either Fortran with added OpenMP or Fortran with HIP bindings and extracted C kernels (using \href{https://github.com/ROCmSoftwarePlatform/hipfort}{hipfort}). The covered functionality of GPUFORT is driven by use-case requirements, the last commit is two years old. Support for OpenACC Fortran is also available by the community through \href{https://gcc.gnu.org/onlinedocs/gfortran/OpenACC.html}{GCC (\texttt{gfortran})} and upcoming in \href{https://ieeexplore.ieee.org/document/9651310}{LLVM (Flacc)}. Also the \href{https://cpe.ext.hpe.com/docs/cce/man7/intro_openacc.7.html}{HPE Cray Programming Environment supports OpenACC Fortran} on AMD GPUs. In addition, the \href{https://github.com/intel/intel-application-migration-tool-for-openacc-to-openmp}{translator tool to convert OpenACC source to OpenMP source by Intel} can be used.  \mycite{gpufort,gccopenacc,flaccieee} 

\mymarker{amdopenmpc}{AMD}{OpenMP}{C}{['fullok', 'nonvendorok']} AMD offers \href{https://github.com/ROCm-Developer-Tools/aomp}{AOMP}, a dedicated, Clang-based compiler for using OpenMP C/C++ on AMD GPUs (\emph{offloading}). AOMP is usually shipped with ROCm. The compiler \href{https://www.exascaleproject.org/wp-content/uploads/2022/02/Elwasif-ECP-sollve_vv_final.pdf}{supports most OpenMP 4.5 and some OpenMP 5.0 features}. Since the compiler is Clang-based, the usual Clang compiler options apply (\texttt{-fopenmp} to enable OpenMP parsing, and others). Also in the upstream Clang compiler, \href{https://clang.llvm.org/docs/OffloadingDesign.html}{AMD GPUs can be targeted through OpenMP}; as outlined for NVIDIA GPUs, the support for OpenMP 5.0 is nearly complete, and support for OpenMP 5.1/5.2 is comprehensive. In addition, the \href{https://cpe.ext.hpe.com/docs/cce/man7/intro_openmp.7.html}{HPE Cray Programming Environment} supports OpenMP on AMD GPUs.  \mycite{aomp,ecpopenmpbof,hpepe} 

\mymarker{amdopenmpfortran}{AMD}{OpenMP}{F}{fullok} Through \href{https://github.com/ROCm-Developer-Tools/aomp}{AOMP}, AMD supports OpenMP offloading to AMD GPUs in Fortran, using the \texttt{flang} executable and Clang-typical compiler options (foremost \texttt{-fopenmp}). Support for AMD GPUs is also available through the \href{https://cpe.ext.hpe.com/docs/cce/man7/intro_openmp.7.html}{HPE Cray Programming Environment}.  \mycite{aomp,hpepe} 

\mymarker{amdstandardc}{AMD}{Standard}{C}{['nonvendorok', 'prettyok', 'somesupport']} AMD does not yet provide production-grade support for Standard-language parallelism in C++ for their GPUs. Currently under development is \href{https://github.com/ROCmSoftwarePlatform/roc-stdpar}{\emph{roc-stdpar}} (ROCm Standard Parallelism Runtime Implementation), which aims to supply pSTL algorithms on the GPU and \href{https://discourse.llvm.org/t/rfc-adding-c-parallel-algorithm-offload-support-to-clang-llvm/72159}{merge the implementation with upstream LLVM}. Support for GPU-parallel algorithms is enabled with \texttt{-stdpar}. An \href{https://discourse.llvm.org/t/rfc-openmp-offloading-backend-for-c-parallel-algorithms/73468}{alternative proposal in the LLVM} community aims to support the pSTL via an OpenMP backend. Also Open SYCL \href{https://github.com/OpenSYCL/OpenSYCL/pull/1088}{is in the process of creating support for C++ parallel algorithms} via a \texttt{-\/-hipsycl-stdpar} switch. By using Open SYCL\textquotesingle s backends, also AMD GPUs are supported. Intel provides the Open Source \href{https://github.com/oneapi-src/oneDPL}{oneDPL} (\emph{oneAPI DPC++ Library}) which \href{https://oneapi-src.github.io/oneDPL/parallel_api_main.html}{implements pSTL algorithms} through the DPC++ compiler (see also \emph{C++ Standard Parallelism for Intel GPUs}). DPC++ has \href{https://intel.github.io/llvm-docs/GetStartedGuide.html\#build-dpc-toolchain-with-support-for-hip-amd}{experimental support for AMD GPUs}.  \mycite{rocstdpar,opensyclproceedings,onedpl} 

\mymarker{amdstandardfortran}{AMD}{Standard}{F}{nope} There is no (known) way to launch Standard-based parallel algorithms in Fortran on AMD GPUs. 

\mymarker{amdkokkosc}{AMD}{Kokkos}{C}{nonvendorok} \href{https://github.com/kokkos/kokkos}{Kokkos} supports AMD GPUs in C++ mainly through the HIP/ROCm backend. Also, an OpenMP offloading backend is available.  \mycite{kokkos} 

\mymarker{amdalpakac}{AMD}{ALPAKA}{C}{nonvendorok} \href{https://github.com/alpaka-group/alpaka}{Alpaka} supports AMD GPUs in C++ through HIP or through an OpenMP backend.  \mycite{alpaka} 

\mymarker{amdpython}{AMD}{etc}{Python}{somesupport} AMD does not officially support GPU programming with Python, but third-party solutions are available. \href{https://docs.cupy.dev/en/latest/install.html\#using-cupy-on-amd-gpu-experimental}{CuPy} experimentally supports AMD GPUs/ROCm. The package can be found on PyPI as \texttt{cupy-rocm-5-0}. Numba once had \href{https://numba.pydata.org/numba-doc/latest/roc/index.html}{support for AMD GPUs}, but it is \href{https://numba.readthedocs.io/en/stable/release-notes.html\#version-0-54-0-19-august-2021}{not maintained anymore}. Low-level bindings from Python to HIP exist, for example \href{https://github.com/jatinx/PyHIP}{PyHIP} (available as \texttt{pyhip-interface} on PyPI). Bindings to OpenCL also exist (\href{https://documen.tician.de/pyopencl/}{PyOpenCL}).  \mycite{cudapython} 

\mymarker{intelcudac}{Intel}{CUDA}{C}{['indirectok', 'nonvendorok']} Intel itself does not support CUDA C/C++ on their GPUs. They offer \href{https://github.com/oneapi-src/SYCLomatic}{SYCLomatic}, though, an Open Source tool to translate CUDA code to SYCL code, allowing it to run on Intel GPUs. The commercial variant of SYCLomatic is called the \href{https://www.intel.com/content/www/us/en/developer/tools/oneapi/dpc-compatibility-tool.html}{DPC++ Compatibility Tool} and bundled with oneAPI toolkit. The community project \href{https://github.com/CHIP-SPV/chipStar}{chipStar} (previously called CHIP-SPV, recently released a 1.0 version) allows to target Intel GPUs from CUDA C/C++ code by using the CUDA support in Clang. chipStar delivers a \href{https://github.com/CHIP-SPV/chipStar/blob/main/docs/Using.md\#compiling-cuda-application-directly-with-chipstar}{Clang-wrapper, \texttt{cuspv}}, which replaces calls to \texttt{nvcc}. Also \href{https://github.com/vosen/ZLUDA}{ZLUDA} exists, which implements CUDA support for Intel GPUs; it is not maintained anymore, though.  \mycite{syclomatic,chipstar,oneapi} 

\mymarker{intelcudafortran}{Intel}{CUDA}{F}{nope} No direct support exists for CUDA Fortran on Intel GPUs. A simple example to bind SYCL to a (CUDA) Fortran program (via ISO C BINDING) can be \href{https://github.com/codeplaysoftware/SYCL-For-CUDA-Examples/tree/master/examples/fortran_interface}{found on GitHub}. 

\mymarker{intelhipc}{Intel}{HIP}{C}{nonvendorok} No native support for HIP C++ on Intel GPUs exists. The Open Source third-party project \href{https://github.com/CHIP-SPV/chipStar}{chipStar} (previously called CHIP-SPV), though, supports \href{https://github.com/CHIP-SPV/chipStar/blob/main/docs/Using.md\#compiling-a-hip-application-using-chipstar}{HIP on Intel GPUs} by mapping it to OpenCL or Intel\textquotesingle s Level Zero runtime. The compiler uses an LLVM-based toolchain and relies on its HIP and SPIR-V functionality.  \mycite{chipstar} 

\mymarker{intelhipfortran}{Intel}{HIP}{F}{nope} HIP for Fortran does not exist, and also no translation efforts for Intel GPUs. 

\mymarker{intelsyclc}{Intel}{SYCL}{C}{fullok} \href{https://www.khronos.org/sycl/}{SYCL} is a C++17-based standard and selected by Intel as the prime programming model for Intel GPUs. Intel implements SYCL support for their GPUs \href{https://github.com/intel/llvm}{via DPC++}, an LLVM-based compiler toolchain. Currently, Intel maintains an own fork of LLVM, but \href{https://lists.llvm.org/pipermail/cfe-dev/2019-January/060811.html}{plans to upstream the changes} to the main LLVM repository. Based on DPC++, Intel releases a \href{https://www.intel.com/content/www/us/en/developer/tools/oneapi/dpc-compiler.html}{commercial \emph{Intel oneAPI DPC++} compiler} as part of the \href{https://www.intel.com/content/www/us/en/developer/tools/oneapi/toolkits.html}{oneAPI toolkit}. The third-party project Open SYCL also supports Intel GPUs, by leveraging/creating LLVM support (either SPIR-V or Level Zero). A previous solution for targeting Intel GPUs from SYCL was \href{https://developer.codeplay.com/products/computecpp/ce/home/}{ComputeCpp of CodePlay}. The project became unsupported in September 2023 (in favor of implementations to the DPC++ project).  \mycite{intelllvm,oneapi,opensyclproceedings} 

\mymarker{intelopenaccc}{Intel}{OpenACC}{C}{somesupport} No direct support for OpenACC C/C++ is available for Intel GPUs. Intel offers a Python-based tool to translate source files with OpenACC C/C++ to OpenMP C/C++, the \href{https://github.com/intel/intel-application-migration-tool-for-openacc-to-openmp}{\emph{Application Migration Tool for OpenACC to OpenMP API}}.  \mycite{acc2mp} 

\mymarker{intelopenaccfortran}{Intel}{OpenACC}{F}{somesupport} Also for OpenACC Fortran, no direct support is available for Intel GPUs. Intel\textquotesingle s \href{https://github.com/intel/intel-application-migration-tool-for-openacc-to-openmp}{source-to-source translation tool from OpenACC to OpenMP} also supports Fortran, though.  \mycite{acc2mp} 

\mymarker{intelopenmpc}{Intel}{OpenMP}{C}{fullok} OpenMP is a second key programming model for Intel GPUs and \href{https://www.intel.com/content/www/us/en/develop/documentation/get-started-with-cpp-fortran-compiler-openmp/top.html}{well-supported by Intel}. For C++, the support is built into the commercial version of DPC++/C++, \emph{Intel oneAPI DPC++/C++}. All \href{https://www.intel.com/content/www/us/en/developer/articles/technical/openmp-features-and-extensions-supported-in-icx.html}{OpenMP 4.5 and most OpenMP 5.0 and 5.1 features are supported}. OpenMP can be enabled through the \texttt{-qopenmp} compiler option of \texttt{icpx}; a suitable offloading target can be given via \texttt{-fopenmp-targets=spir64}.  \mycite{oneapi} 

\mymarker{intelopenmpfortran}{Intel}{OpenMP}{F}{fullok} OpenMP in Fortran is Intel\textquotesingle s main selected route to bring Fortran applications to their GPUs. OpenMP offloading in Fortran is supported through \href{https://www.intel.com/content/www/us/en/docs/fortran-compiler/developer-guide-reference/2023-2/overview.html}{Intel\textquotesingle s Fortran Compiler \texttt{ifx}} (the new LLVM-based version, not the \emph{Fortran Compiler Classic}), part of the oneAPI HPC Toolkit. Similarly to C++, OpenMP offloading can be enabled through a combination of \texttt{-qopenmp} and \texttt{-fopenmp-targets=spir64}.  \mycite{oneapi} 

\mymarker{intelstandardc}{Intel}{Standard}{C}{['fullok', 'prettyok']} Intel supports C++ standard parallelism (\emph{pSTL}) through the Open Source \href{https://oneapi-src.github.io/oneDPL/index.html}{oneDPL} (oneAPI DPC++ Library), also available as part of the oneAPI toolkit. It \href{https://oneapi-src.github.io/oneDPL/parallel_api_main.html}{implements the pSTL} on top of the DPC++ compiler, algorithms, data structures, and policies live in the \texttt{oneapi::dpl::} namespace. In addition, \href{https://github.com/OpenSYCL/OpenSYCL/pull/1088}{Open SYCL is current adding support for C++ parallel algorithms}, to be enabled via the \texttt{-\/-hipsycl-stdpar} compiler option.  \mycite{onedpl} 

\mymarker{intelstandardfortran}{Intel}{Standard}{F}{fullok} Standard language parallelism of Fortran is supported by Intel on their GPUs through the Intel Fortran Compiler \texttt{ifx} (the new, LLVM-based compiler, not the \emph{Classic} version), part of the oneAPI HPC toolkit. In the \href{https://www.intel.com/content/www/us/en/developer/articles/release-notes/fortran-compiler-release-notes.html}{oneAPI update 2022.1}, the \href{https://www.intel.com/content/www/us/en/docs/fortran-compiler/developer-guide-reference/2023-2/do-concurrent.html}{\texttt{do\ concurrent} support} was added and extended in further releases. It can be used via the \texttt{-qopenmp} compiler option together with \texttt{-fopenmp-target-do-concurrent} and \texttt{-fopenmp-targets=spir64}.  \mycite{oneapi} 

\mymarker{intelkokkosc}{Intel}{Kokkos}{C}{nonvendorok} No direct support by Intel for Kokkos is available, but \href{https://kokkos.github.io/kokkos-core-wiki/}{Kokkos} supports Intel GPUs through an experimental SYCL backend.  \mycite{kokkos} 

\mymarker{intelalpakac}{Intel}{ALPAKA}{C}{nonvendorok} Since \href{https://github.com/alpaka-group/alpaka/releases/tag/0.9.0}{v.0.9.0}, \href{https://github.com/alpaka-group/alpaka}{Alpaka} contains experimental SYCL support with which Intel GPUs can be targeted. Also, Alpaka can fall back to an OpenMP backend. 

\mymarker{intelpython}{Intel}{etc}{Python}{prettyok} Intel GPUs can be used from Python through three notable packages. First, Intel\textquotesingle s \href{https://github.com/IntelPython/dpctl}{\emph{Data Parallel Control} (dpctl)} implements low-level Python bindings to SYCL functionality. It is available on PyPI as \href{https://pypi.org/project/dpctl/}{\texttt{dpctl}}. Second, a higher level, Intel\textquotesingle s \href{https://github.com/IntelPython/numba-dpex}{\emph{Data-parallel Extension to Numba} (numba-dpex)} supplies an extension to the JIT functionality of Numba to support Intel GPUs. It is available from Anaconda as \href{https://anaconda.org/intel/numba-dpex}{\texttt{numba-dpex}}. Finally, and arguably highest level, Intel\textquotesingle s \href{https://github.com/IntelPython/dpnp}{\emph{Data Parallel Extension for Numpy} (dpnp)} builds up on the Numpy API and extends some functions with Intel GPU support. It is available on PyPI as \href{https://pypi.org/project/dpnp/}{\texttt{dpnp}}, although latest versions appear to be available \href{https://github.com/IntelPython/dpnp/releases}{only on GitHub}.  \mycite{dpctl,numba-dpex,dpnp} 

}

\section{Discussion}
\label{sec:limitations}

While the possible combinations were explained extensively and the given ratings motivated thoroughly, some limitations and caveats exist.

\subsubsection*{Model Selection}

The most prominent limitation is the selection of programming models. CUDA, HIP, SYCL, OpenACC, OpenMP, Standard Parallelism, Kokkos, Alpaka, and \emph{Python} were selected for their prevalence in the HPC community -- and to focus the scope of this work. But of course, there are further programming models available and used in the community.
The most notable exclusion is certainly RAJA~\autocite{raja}. The choice for omitting was made because it is similar in spirit to, albeit not as popular as Kokkos\footnote{Although GitHub \emph{Stars} are inherently a flawed metric, RAJA has about one-third as many stars as Kokkos}. 
OpenCL~\autocite{opencl} is a further important GPU programming model, but it has never gained much traction in the HPC-GPU space, mostly due to the lukewarm support by NVIDIA. 
Other models exist, like HPX~\autocite{hpx} (which is similar to pSTL support, arguably more extensive, but less \emph{standard}) or C++AMP~\autocite{cppamp} (which was deprecated in 2022). 
In principle, also the core of PyTorch~\autocite{libtorch}, libtorch, can be used as a form of programming model. 
No compatibility libraries were included either, like the \emph{libompx} project~\autocite{libompx} which prototypes implementing vendor-agnostic pSTL-like algorithms.

\subsubsection*{Performance Evaluation}

A second important limitation is the lack of evaluation of performance. As shown above, many models exist and can target the various GPU types, partly even through different backends. Assessing the level of support, as done here, is at times already challenging; partly even dedicated test suites exist (for example for OpenACC~\autocite{openaccvvwebsite,openmpvvproceedings} or OpenMP~\autocite{openmpvvwebsite,openmpvvproceedings}). Judging the performance fairly is even more involved, as a representative selection of micro-benchmarks would need to be ported to the models. For sub-sets of the presented models, performance comparisons do exist. For example by \citeauthor{hammondgtc}~\autocite{hammondgtc}, who compares many NVIDIA-GPU-compatible programming models (and even various implementation routes). Frequently, application-specific use-cases are evaluated on two or more models/devices. An example is by \citeauthor{lartpc}~\autocite{lartpc}, comparing performance of a physics simulation between Kokkos, SYCL, and OpenMP. Closest to an performance overview certainly gives the BabelStream project~\autocite{babelstream}, although only for a STREAM-like algorithm; an example of a recent performance-comparing publication is~\autocite{babelstreamresults}.

\subsubsection*{Topicality}

Parts of the field are rapidly evolving. For example, the support for C++ standard parallelism on AMD GPUs made great progress in the past year, and now multiple venues exist. Much of the support is driven by the community, especially for the AMD platform, and it can be hard to assess the current status. In addition, proper documentation sometimes does not exists (yet) and one needs to review the source code. At times, some features are not even advertised in the documentation (like the pSTL support on NVIDIA/AMD GPUs through DPC++). The downside of this evolving field are unmaintained models. For example, it is unclear if GPUFORT~\autocite{gpufort} is still \emph{officially} supported by AMD. This paper can hence only be seen current at the time of submission.

\subsubsection*{Individual Category Discussions}

Of course, some assessments are subject to discussion. For example the \emph{\textbf{some support}} category, which is mostly used for \emph{incomplete} support in \autoref{table:compat}. OpenACC C++ support on NVIDIA GPUs (\ref{openaccc})
was rated complete, while OpenMP C++ support showed some ambivalence (\ref{nvidiaopenmpc}). Here, the assessment was made because NVIDIA is upfront in acknowledging that some features of OpenMP for GPU offloading are still missing. Further ambivalence in rating can be seen for Python on NVIDIA GPUs, were the pick-up of the Open Source community was acknowledged through the added \emph{\textbf{non-vendor support}} category. C++ standard language parallelism at AMD has most ambivalence, a result of the rapidly changing support through multiple venues -- and currently no vendor-supported, advertised solution (which roc-stdpar~\autocite{rocstdpar} might become). The double-rating of CUDA C++ on Intel GPUs honors the research project chipStar~\autocite{chipstar}, besides the CUDA-to-SYCL conversion tool~\autocite{syclomatic} by Intel. Finally, C++ standard parallelism for Intel GPUs has ambivalence, as all pSTL functionality currently resides in a custom namespace.

Although due care has been taken when compiling the presented data, there might still be unexplored venues, or changed status. The presented descriptions of this paper reside in a GitHub repository and are open for collaboration through issues or pull requests~\autocite{gpucompat}.

\section{Conclusion}
\label{sec:conclusion}

This paper presented a methodology to categorize the support of programming models on  HPC GPU devices, assessing the level of support and the provider (vendor or third-party). The results for a number of selected models on GPUs of three vendors (AMD, Intel, NVIDIA) were presented in \autoref{table:compat}, accompanied by extensive descriptions in \autoref{sec:descriptions}. The limitations of the method and some key caveats of the presentation were discussed in \autoref{sec:limitations}.

The support for NVIDIA GPUs can be considered most comprehensive, founded in their long-time prevalence in the field.CUDA is possibly the most famous GPU programming model, and both other vendors (AMD, Intel) provide tools for converting CUDA C/C++ to their \emph{native} model (HIP, SYCL). AMD designed HIP closely to mimic CUDA-like programming and enable it other platforms. And, indeed, NVIDIA and AMD GPUs can be used from the same source code, and recently also Intel GPUs with chipStar. SYCL is an entirely different programming model compared to CUDA or HIP, but it also supports all three GPU platform; either by the work by Intel or the community (Open SYCL). While OpenACC can be used on NVIDIA and AMD GPUs, support for Intel GPUs does not exist. OpenMP, on the other hand, is supported on all three platforms -- and even for both C++ and Fortran. Standard language parallelism appears to be the model with the fastest change at the moment, with multiple new projects in progress for all three platforms. Kokkos and Alpaka both provide higher-level abstractions and support all three platform. Python, a somewhat outlier in the list, is also well-supported by all three platforms.

While the C++ support appears to be well on the way to good compatibility and portability, the situation looks severely different for Fortran. The only natively supported programming model on all three platforms is OpenMP.

A key component in the ecosystem is the LLVM toolchain. The compilers of AMD, Intel, and NVIDIA are all based on LLVM infrastructure and partly take great effort in upstreaming their changes. Notable are also the open licenses attached to many components, even the key ecosystem compilers (AMD, Intel). Through LLVM, many third-party/community projects are enabled, which now add valuable contributions to the ecosystem (for example Open SYCL).

Not assessed in this work was the performance of programming models. It is a hard task, but might become a future venue for extension of the presented material. Of course, the landscape of \autoref{table:compat} evolves swiftly; the progress is tracked in a GitHub repository~\autocite{gpucompat}, open for suggestions.

\begin{acks}
	A previous version of this work was shown in a presentation at a workshop~\autocite{hertennatesm}. As it led to many questions and partly heated discussions, the author decided to create a stand-alone version of the comparison and publish it on GitHub~\autocite{gpucompat} with an accompanying blogpost~\autocite{hertencompatblog}, now utilizing source data in YAML form with conversion to HTML and TeX. As interest continued, the material was updated and significantly extended to become this paper. The goal is a living overview of the evolving field, with snapshots in paper form at regular intervals.

	The author would like to acknowledge the frequent discussions with his colleague of Jülich Supercomputing Centre, trying to understand the level of support and determine also corner-cases and unpublished routes of GPU support.
\end{acks}

\printbibliography
\end{document}